\documentclass[reprint,superscriptaddress,aps,prb,twocolumn,floatfix]{revtex4-2}
\usepackage[utf8]{inputenc}
\usepackage{textcomp} 
\usepackage{setspace}
\usepackage{amsmath}
\usepackage{breqn}
\usepackage{graphicx}
\usepackage{verbatim}

\usepackage{amsfonts}
\usepackage{amssymb}
\usepackage{xcolor}
\usepackage{soul}
\usepackage{tabularx}
\setcitestyle{super}
\usepackage{float}
\usepackage[colorlinks,linkcolor=blue,anchorcolor=blue,citecolor=blue,urlcolor=black]{hyperref}
\DeclareGraphicsExtensions{.pdf,.eps,.png,.jpg,.mps}
\usepackage{booktabs}
\usepackage{makecell}
\usepackage{threeparttable}

\usepackage{changepage}

\begin{document}

\title{Integrated photonic ultrawideband real-time spectrum sensing for 6G wireless networks}

\author{Yuansheng Tao$^{1,\dagger,*}$, Hanke Feng$^{1,\dagger,*}$, Yuan Fang$^{2}$, Xiangzhi Xie$^{1}$, Yuansong Zeng$^{1}$, Yifan Wu$^{1}$, Tong Ge$^{1}$, Yiwen Zhang$^{1}$, Zhaoxi Chen$^{1}$, Zihan Tao$^{3}$, Jie Xu$^{4}$, Haowen Shu$^{3}$, Xingjun Wang$^{3}$, Xianghao Yu$^{2}$, and Cheng Wang$^{1,*}$ \\
\vspace{3pt}
$^1$Department of Electrical Engineering \& State Key Laboratory of Terahertz and Millimeter Waves, City University of Hong Kong, Kowloon, China.\\
$^2$Department of Electrical Engineering, City University of Hong Kong, Kowloon, China \\
$^3$State Key Laboratory of Photonics and Communications, School of Electronics, Peking University, Beijing, China \\
$^4$School of Science and Engineering \& Guangdong Provincial Key Laboratory of Future Networks of Intelligence, Chinese University of Hong Kong (Shenzhen), Guangdong 518172, China \\
$^{\dagger}$These authors contributed equally to this work \\
\vspace{3pt}
Corresponding authors: yuanstao@um.cityu.edu.hk, hankefeng2@cityu.edu.hk, cwang257@cityu.edu.hk}

\begin{abstract}
\begin{adjustwidth}{-1.2cm}{0.6cm}
\fontsize{10.5pt}{12pt}\selectfont
\textbf{Abstract} \vspace{3pt}\\
\fontsize{10.2pt}{12pt}\selectfont
The sixth generation (6G) wireless networks require dynamic spectrum management to optimize the utilization of scarce spectral resources and support emerging integrated sensing and communication (ISAC) applications. This necessitates real-time spectrum sensing (RT-SS) capability with ultrawide measurement range, compact size, and low latency. Conventional electronic RT-SS solutions face critical challenges in operating across the millimeter-wave and sub-terahertz bands, which are essential spectra for 6G wireless. While photonic RT-SS has the potential to surpass this limitation, the current implementations feature limited bandwidths below 50 GHz and mostly rely on bulky dispersive fibers with high latency. Here, we address these challenges by developing an integrated photonic RT-SS system capable of ultrabroadband measurement from microwave to sub-terahertz bands, covering the full spectrum for 6G wireless. The photonic RT-SS circuit integrates a broadband electro-optic (EO) modulator for unknown signal loading, an EO tunable microring filter bank for high-speed parallel frequency-to-time mapping, as well as an EO comb for precise channel frequency referencing, all realized on a single thin-film lithium niobate chip. We achieve an unprecedented spectral measurement range beyond 120 GHz at a low latency of less than 100 ns. To validate the effectiveness of our photonic RT-SS system in future 6G scenes, we further propose a heuristic spectro-temporal resource allocation algorithm and conduct a proof-of-concept ISAC demonstration, where a radar adaptively access RT-SS-informed spectrally underutilized regions for high-quality target sensing under dynamic communication interferences. Our work presents a compact and cost-effective solution for efficient spectrum sharing and dynamic management in future 6G ISAC networks.

\end{adjustwidth}

\end{abstract}

\maketitle

\begin{figure*}[ht]
\centering
\includegraphics[width = 18cm]{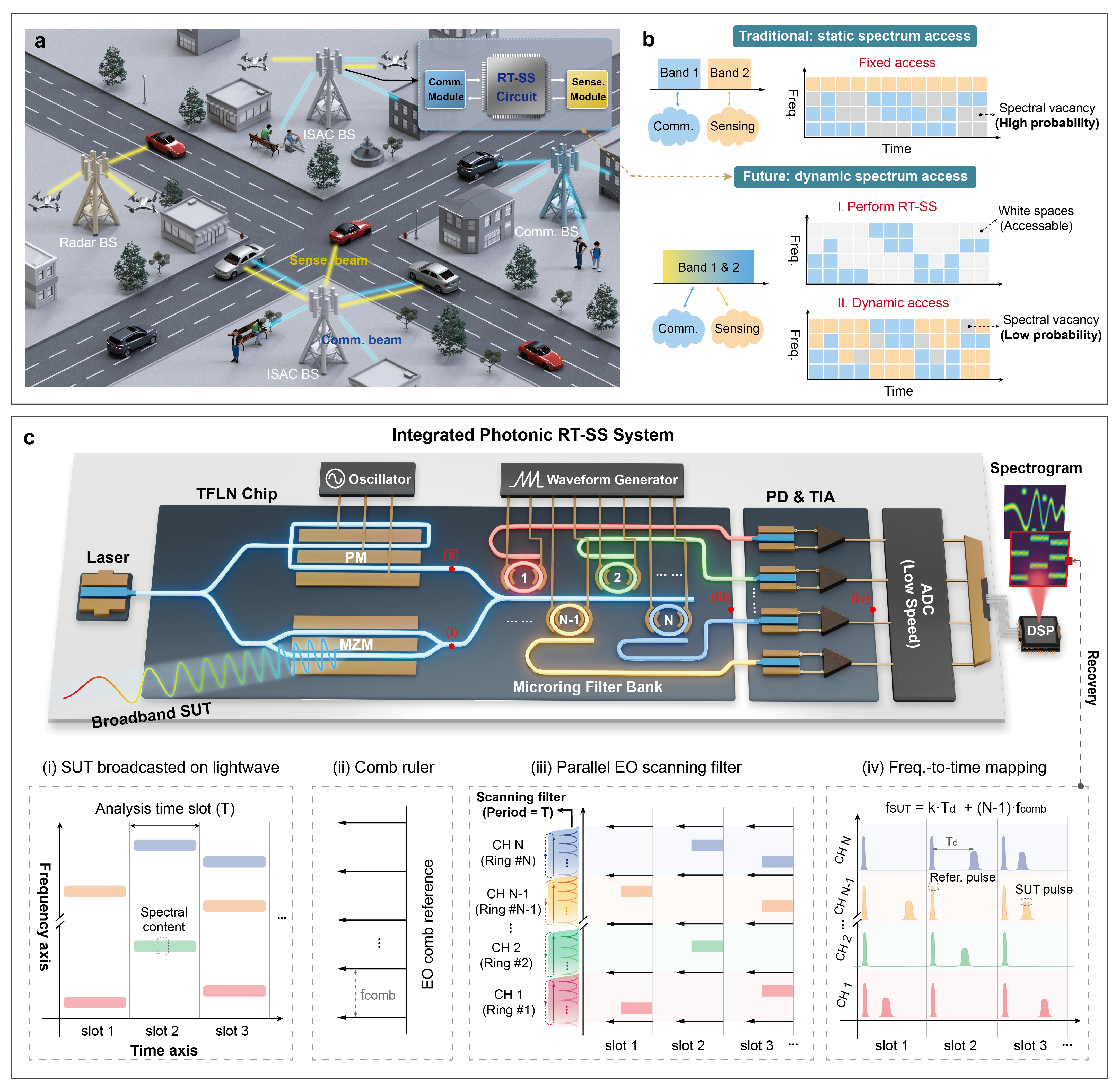}
\caption{\textbf{Integrated photonic real-time spectrum sensing (RT-SS) for 6G integrated sensing and communications (ISAC).} \textbf{a}, An envisioned ISAC scene facilitated by the photonic RT-SS module for dynamic spectrum resource allocation between radar sensing and communication channels. \textbf{b}, Schematic comparison of traditional static spectrum access (SSA) and future dynamics spectrum access (DSA) spectrum management schemes. Unlike the SSA scheme where the spectrum resources are rigidly allocated to sensing and communication users by government regulations, the DSA scheme dynamically allocates the entire spectrum resources based on real-time spectrum usage information. For instance, sensing acts as secondary users that adaptively access the underutilized spectral white spaces while avoiding interferences from communication (primary user). \textbf{c}, Schematic diagram and operation principles of the integrated photonic RT-SS system. Time-varying spectral information of incoming signal under test (SUT) is broadcasted onto an optical carrier by a Mach-Zehnder modulator (i). The modulated optical signal is subsequently processed in parallel by an electro-optic (EO) microring filter bank, where each microring is biased at slightly different resonance wavelength and responsible for monitoring a specific spectral range (color coded in insets). During the analysis time slot, denoted as T, each microring filter is EO scanned across its responsible spectral range (iii), mapping the respective spectral information into time-domain signals (iv). To ensure accurate frequency calibration across different channels, an EO comb is generated from the same optical carrier by an on-chip phase modulator (ii). This frequency comb ruler is combined with the SUT path and also mapped to the time domain. The final output temporal waveforms after photodetection in each analysis time slot consists of a reference pulse and the SUT pulses, which are used to recover the unknown spectral contents of the original SUT. }
\label{fig1}
\end{figure*}

\vspace{3pt}
\noindent
\textbf{Introduction} \\
Radio-frequency (RF) spectrum plays a fundamental role in the development of radar sensing and wireless communications over past decades \cite{1,2}. Recently, integrated sensing and communication (ISAC) has emerged as a pivotal concept for the upcoming sixth generation (6G) wireless networks \cite{3}, where sensing and communication functions are expected to be unified in the same system over shared spectral bands to support various environment-aware applications \cite{4,5}, as envisioned in Fig. \ref{fig1}a. In this case, the inherent competition for spectrum access between these functionalities will lead to severe spectrum congestion problems \cite{6,7}. To address this challenge, a paradigm shift in RF spectrum management is required to transform conventional static spectrum access (SSA) to more advanced dynamic spectrum access (DSA), enabling cognitive radios with more efficient spectrum sharing and utilization \cite{8,9}, as illustrated in Fig. \ref{fig1}b. Specifically, radar sensing (e.g., as secondary user) could coexist with communication (primary user) over the entire spectrum, through dynamically recognizing and accessing underutilized spectral bands. As such, the spectrum allocation could more flexibly adapt to real-time communication/sensing bandwidth demands, avoiding spectrum vacancy and enhancing spectrum utilization efficiency.

Real-time spectrum sensing (RT-SS) is a critical enabler of this dynamic RF spectrum access vision in 6G ISAC networks \cite{10, 11}, where RT-SS modules are installed within ISAC base stations (inset of Fig. \ref{fig1}a) to monitor real-time spectrum usage of the electromagnetic environment, thereby facilitating adaptive spectrum allocation for radar sensing and communication services. To exploit the full prospect of 6G ISAC \cite{12}, the corresponding RT-SS should simultaneously feature ultrabroad bandwidth ($>$100 GHz), low processing latency, and high temporal resolution (both at nanosecond-level). This is essential for utilizing all spectrum bands, from microwave to millimeter-wave (mmWave) and even terahertz (THz), to deliver the anticipated Tbps-rate data communications \cite{13} and millimeter-level radar sensing \cite{14} in a spectrum-agile wireless environment in the 6G era. Moreover, RT-SS systems should be implemented on integrated platforms to significantly reduce size, weight, and power consumption (SWaP) to accommodate the widespread deployment of ISAC base stations for massive and ubiquitous user connections \cite{15}.

Conventionally, RT-SS is realized through electronic-based approaches \cite{16}, such as digital sampling \cite{17}, sweeping local oscillators \cite{18}, and dispersive phase shifters \cite{19, 20}. However, owing to the intrinsic bandwidth limitations of electronic devices, all these electronic RT-SS implementations operate with measurement bandwidths below 20 GHz. Photonic RT-SS provides a promising alternative to overcome the bandwidth limitations of their electronic counterparts \cite{21,22}. Typical photonic RT-SS systems utilize chromatic dispersion in optical fibers to perform the frequency to time mapping (FTTM) operation for spectral information recovery \cite{23, 24, 25, 26, 27, 28}, which support relatively broad analysis bandwidths up to 46 GHz \cite{28}. However, the long dispersive fibers involved in the systems usually introduce large processing latency ($>$100 $\mu$s) \cite{23,24}, fundamentally hindering the real-time analysis ability. To reduce the latency, linearly chirped fiber Bragg gratings are employed to achieve large dispersion in a short physical length \cite{28}. Alternatively, various dispersion-free photonic RT-SS architectures have been proposed, such as passive fiber loops \cite{29} and frequency shifted feedback lasers \cite{30} leveraging the Talbot effect, as well as the spectral gating approaches \cite{31,32,33,34}. However, these photonic implementations still rely on discrete optoelectronic devices with significant SWaP disadvantages and are limited to analysis bandwidths of less than 50 GHz \cite{34}. Recently, integrated photonic RT-SS systems have been attempted on silicon photonic platform, through constructing frequency-scanning optical filters to perform FTTM-based spectrum reconstruction \cite{35,36,37}. However, the achieved temporal resolution is at millisecond timescale, inherently restricted by the tuning speed of thermal heaters, which is insufficient to capture fast-changing spectral environment in the 6G wireless. Meanwhile, their measurement bandwidths are usually lower than 35 GHz limited by the electro-optic (EO) bandwidth of silicon modulators \cite{38}, far from that needed for directly analyzing mmWave/THz spectrum. In short, realizing a photonic RT-SS system that can simultaneously meet the bandwidth, latency, temporal resolution, and integration requirements for 6G ISAC has remained elusive.

In this Article, we fulfill this promise by demonstrating a photonic chip-based, ultrabroadband RT-SS system based on thin-film lithium niobate (TFLN) platform. In recent years, the TFLN platform has demonstrated large-bandwidth EO modulators \cite{39,40,41} for RF-to-optic broadcasting, high-efficiency EO combs \cite{42,43,44} for spectral parallelization, and low-loss, fast-tunable microring resonators \cite{45,46} for agile spectral manipulation. Leveraging these exceptional building blocks and wafer-scale scalability \cite{47}, we address the performance limitations of previous works, particularly unlocking the bandwidth potential of photonic RT-SS implementation. In our work, EO-driven microrings are utilized as high-speed frequency-scanning filters to perform the FTTM operation with nanosecond temporal resolution. To overcome the limited EO tuning range of a single microring, a channelized measurement architecture referenced by an EO comb ruler is proposed, where parallelly measured spectral slices can be seamlessly stitched together to recover the entire spectrum of ultrabroadband signals. Thanks to the collective benefits from TFLN platform and the channelized methodology, the photonic RT-SS system achieves an unprecedented spectral analysis range up to 120 GHz at a low latency of $<$100 ns, along with an advanced frequency resolution of 350 MHz and a temporal resolution of 100 ns. To showcase its real-world applicability, we further apply the photonic RT-SS to a proof-of-concept ISAC scenario, where the communication and radar sensing dynamically share the same spectral resource. We propose and develop a spectro-temporal resource allocation algorithm that can utilize the dynamic spectral usage information provided by the RT-SS to adaptively find the optimal accessible frequency bands with significantly suppressed mutual interferences. As a result, our ISAC system achieves an 8.8-dB enhancement of signal-to-interference-plus-noise ratio (SINR) for radar ranging under dynamic spectral interferences from communication channels. Our work marks a significant step toward dynamic spectrum management in 6G ISAC wireless networks.

\begin{figure*}[ht]
\centering
\includegraphics[width = 18cm]{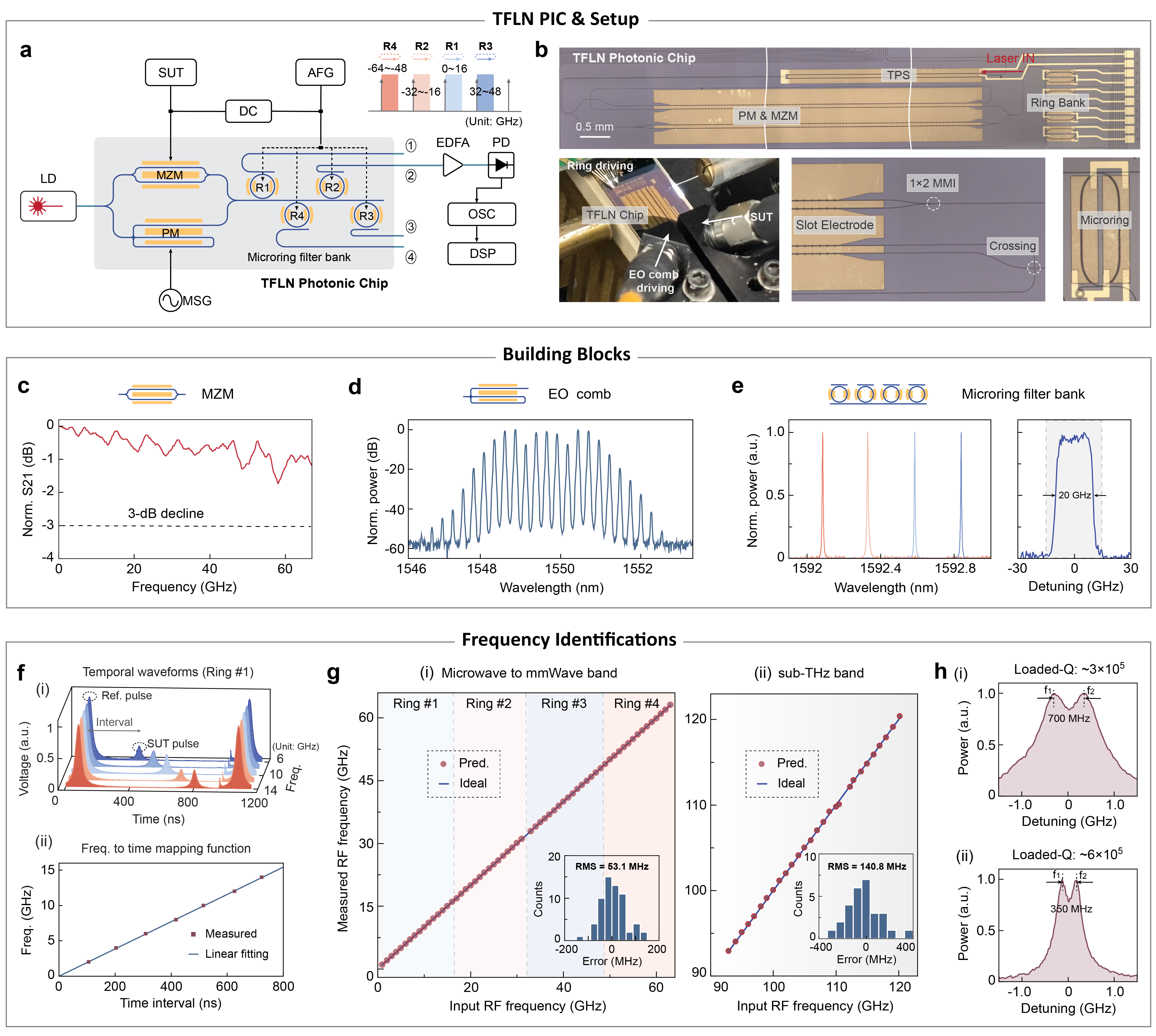}
\caption{\textbf{Device characterizations and ultrawideband high-precision frequency measurement.} \textbf{a}, Experimental setup of the TFLN photonic chip-based RT-SS system. SUT, signal under test; MSG, microwave source generator; AFG, arbitrary function generator; EDFA, erbium-doped fiber amplifier; PD, photodetector; OSC, oscilloscope; DSP, digital signal processing. \textbf{b}, Microscope images of the TFLN photonic RT-SS chip and test board. \textbf{c}, Electro-optic (EO) S21 responses of the on-chip Mach-Zehnder modulator. \textbf{d}, Optical spectrum of the on-chip EO frequency comb, with a repetition rate of 32 GHz. \textbf{e}, Left: static drop-port transmission spectra of the microring filter bank with equal spacing of 32 GHz. Right: optical transmission spectrum of one microring filter actively driven by a 20 Vpp ramp waveform, indicating a 20-GHz EO frequency sweeping range. \textbf{f}, (i) Measured electrical temporal waveform traces in channel 1, corresponding to calibration input frequencies from 6 to 14 GHz at a step of 2 GHz. (ii) Linear fitting of the RF frequencies versus time intervals between reference pulses and SUT pulses. \textbf{g}, Measured RF frequencies versus input frequencies of single-tone SUTs from 0 to 62 GHz (i) and from 90 to 120 GHz (ii). Insets: Statistical histograms of the frequency measurement errors. \textbf{h}, Measured temporal waveforms of dual-tone input SUTs for evaluating the frequency resolutions, using microring filters with loaded $Q$ factors of $3\times10^{5}$ (i) and $6\times10^{5}$ (ii), respectively.}
\label{fig2}
\end{figure*}

\vspace{3pt}
\noindent
\textbf{Results} \\
\textbf{Principles of the integrated photonic RT-SS} \\
\noindent Fig. \ref{fig1}c illustrates the schematic and working principle of the proposed TFLN photonic RT-SS system. A continuous wave (c.w.) optical carrier is injected into the TFLN chip and subsequently split into two optical paths. The bottom path connects to an MZM, which is used to transfer broadband incoming RF signal under test (SUT) to the optical domain, as shown in inset (i). The top path connects to a recycling phase modulator (PM), which is driven by an RF oscillator to generate an EO frequency comb. The EO comb lines serve as frequency references in the following channelized spectral analysis process [inset (ii)]. The optically loaded SUT and comb references are subsequently combined and launched into a microring filter bank, where each microring is biased at a slightly different center wavelength and is responsible for monitoring a certain spectral slice [inset (iii)]. The microrings are synchronously EO modulated by periodic ramp voltage waveforms to implement frequency-sweep filtering.

Within each spectral channel, we align the microring resonance and one reference comb line such that during one scanning period (denoted as T), the microring first sweeps across the comb line and then the SUT sidebands, resulting in respective time-domain electrical pulses after photodetection [inset (iv)]. In this way, a linear FTTM relationship with absolute frequency reference is established, which ensures precise measurements with robustness against laser wavelength fluctuations and microring resonance variations. By measuring the time intervals between the reference pulses and the SUT pulses using low-speed sampling electrical circuits, the unknown spectral components can be identified. Thanks to the Pockels effect of the TFLN platform, the scanning period (T) of microring filters can be sufficiently short down to nanosecond level. This allows us to regard time-varying SUT as stationary during one filter scanning duration, leading to the implementation of the FTTM (Fourier transform) process in real time. By continuously recording the output waveforms over many analysis time slots, the time-varying frequency information of the entire SUT, namely its spectrogram, can be captured. Finally, the spectrograms measured in different spectral channels can be seamlessly stitched together with the help of EO comb reference, completing the RT-SS process of the full input SUT. Our parallel FTTM scheme fundamentally differs from previously reported TFLN RF frequency measurement systems based on frequency-to-power mapping schemes \cite{48,49}. Such schemes only recognize a single dominating frequency component within a microsecond-level analysis time window and cannot fulfill the complex spectrum reconstruction requirements in 6G ISAC networks. A more detailed analysis of the proposed photonic RT-SS principles is provided in Supplementary Note I.

\vspace{3pt}
\noindent\textbf{Experimental setup and device characterizations}\\
\noindent The TFLN photonic chip-based RT-SS system with a four-channel microring filter bank is experimentally demonstrated using a setup shown in Fig. \ref{fig2}a. In our actual experiment, we make use of both upper and lower optical modulation sidebands in an interleaved manner, as schematically shown in the inset of Fig. \ref{fig2}a. Specifically, we set the EO comb spacing (32 GHz) to twice the scanning spectral range of each microring filter (16 GHz). Microring 1 and microring 3 scan the upper sidebands, referenced by the carrier and +1 order comb line, respectively, whereas microring 2 and microring 4 scan the lower sidebands, referenced by the -1 and -2 order comb lines. Compared to the simplified schematic in Fig. \ref{fig1}c, this interleaved configuration requires only half the number of comb lines to achieve the same RF spectral analysis bandwidth. It also avoids spectral measurement gaps between adjacent scanning ranges of microring filters, ensuring seamless spectral coverage across channels.

The integrated photonic chip is fabricated based on a 4-inch wafer-scale TFLN technology (see Methods), as displayed in Fig. \ref{fig2}b. The TFLN photonic chip consists of several element devices, including a tunable optical power splitter, an MZM, a recycling PM and a four-channel microring filter bank. Before investigating the system-level functionalities, we first characterize the performances of these on-chip building blocks. The on-chip MZM demonstrates a large EO bandwidth significantly exceeding 67 GHz (Fig. \ref{fig2}c) and a low half-wave voltage of 2.65 V. These performances significantly outperform previous works using bulky LN \cite{23,28} or silicon photonic modulators \cite{37}, thereby enabling the faithful broadcasting of broadband SUT spectrum into the optical domain with high conversion efficiency. The recycling PM is designed with optical delay lines to efficiently produce EO combs at the desired repetition rate of 32 GHz, as shown in Fig. \ref{fig2}d. A total of 15 comb lines are generated with a 20-dB optical bandwidth of 3.84 nm. The microring filters are designed with free spectral ranges (FSR) of 125 GHz, which in principle supports an unambiguous RF frequency identification range up to half the FSR, i.e. 62.5 GHz. The measurement range can be extended to a full FSR using carrier-suppressed single-sideband modulation in future optimizations. By applying a 20 V peak-to-peak ramp waveform to each microring filter, an EO scanning range (3-dB decline) of approximately 20 GHz is achieved (Methods), as shown in Fig. \ref{fig2}e (right). Based on these evaluated device performances, in the subsequent photonic RT-SS experiments, the spectral slices of the entire SUT are set to 0-16 GHz, 16-32 GHz, 32-48 GHz, and 48-64 GHz for the four microring filter channels, respectively. The 16-GHz slicing width is slightly narrower than the full microring scanning range to allow partial overlaps between adjacent analysis channels for inter-band spectrum stitching. Correspondingly, the four microring filters are initially biased at an equal frequency spacing of 32 GHz, as shown in Fig. \ref{fig2}e (left).

\vspace{3pt}
\noindent\textbf{High-precision RF frequency identification}\\
\noindent To evaluate the system-level performance of the TFLN photonic RT-SS system, we first validate the RF frequency identification capabilities by applying stationary single-tone and multiple-tone microwave SUTs. Here the periodic scan frequency of the microring filters is set to 1 MHz. We calibrate the system by inputting single-tone RF signals with predefined frequencies at 2 GHz steps. The calibration allows us to obtain the FTTM functions for the four spectral analysis channels, which will be used to estimate unknown SUT frequencies in subsequent experiments. Example original temporal waveforms obtained for channel 1 during the calibration process are shown in Fig. \ref{fig2}f (i). The time intervals between the reference pulses and signal pulses are extracted and linearly fitted to the corresponding input RF frequencies, as shown in Fig. \ref{fig2}f (ii). The excellent FTTM linearity is achieved due to the intrinsically linear Pockels effect of lithium niobate, which ensures precise measurements with minimal signal post-processing. The calibration results of other channels are provided in Supplementary Note II, which all show similarly good linear relationships. The calibrated system is then used to identify unknown single-tone frequencies from 0 to 62 GHz with a step of 1 GHz. The measurement results are shown in Fig. \ref{fig2}g (i), achieving a root mean square (RMS) frequency measurement error of 53.1 MHz. The data point at 32 GHz frequency is absent since it is spectrally occupied by a reference comb line in our current setup.

Leveraging the ultrabroad bandwidth of TFLN modulators, we show that the frequency measurement range can be further extended to sub-THz band up to 120 GHz. In this experiment, we apply sub-THz SUT to the on-chip MZM using a 100-GHz single-port RF probe, whereas the EO comb reference is replaced by an external tunable laser source due to the lack of a dual-port RF probe with sufficient operating bandwidth (see Supplementary Note III for more details). Fig. \ref{fig2}g (ii) presents the frequency identification results within the 90-120 GHz range using our TFLN photonic RT-SS system, with an RMS frequency measurement error of 140.8 MHz. The reduced frequency measurement precision is primarily attributed to random relative wavelength fluctuations, as the reference laser and the carrier laser used for broadcasting the SUT are independently free-running. This issue could be addressed in the future by using broadband dual-port RF probes or high-frequency RF packaging. The increase in measurement error also partly results from weaker optical modulation sidebands at sub-THz frequencies owing to the higher electrode loss of on-chip MZM, which leads to a degradation of signal-to-noise ratio (SNR).

To evaluate the frequency resolution of the photonic RT-SS system, two-tone SUTs with small frequency spacing are generated as input. Here, we define the frequency resolution criterion as: if the center minima between two closely-spaced SUT pulses is more than 1 dB lower than the peak values, we consider the two pulses (or the two corresponding frequencies) resolvable. Based on this resolution criterion, the measured frequency resolution for the 125-GHz microrings is approximately 700 MHz, as shown in Fig. \ref{fig2}h (i). The frequency resolution in our scan-filter based FTTM scheme is primarily determined by the Lorentzian-shaped bandpass filter linewidth of the microring \cite{35}. The obtained 700 MHz resolution aligns well with the extracted loaded quality ($Q$) factor of $3\times10^{5}$ from the drop-port transmission spectrum for the microring. The resolution can be further improved by using a microring filter with a higher $Q$ factor. Using a 70-GHz FSR microring with a loaded $Q$ of $6\times10^{5}$, we obtain a frequency resolution of 350 MHz, as shown in Fig. \ref{fig2}h (ii). It is worth noting that the frequency resolution could further degrade from the filter linewidth if the microring is swept across its resonance with a duration shorter than the cavity photon lifetime. This phenomenon will be discussed with more details in the next section. The 1 $\mu$s scanning period used in the experiments here corresponds to a resonance sweeping duration of about 31.5 ns (see Supplementary Note IV), which is significantly longer than the photon lifetime ($\sim$0.25 ns) and therefore will not lead to degradation of the frequency resolution.

\begin{figure*}[ht]
\centering
\includegraphics[width = 18cm]{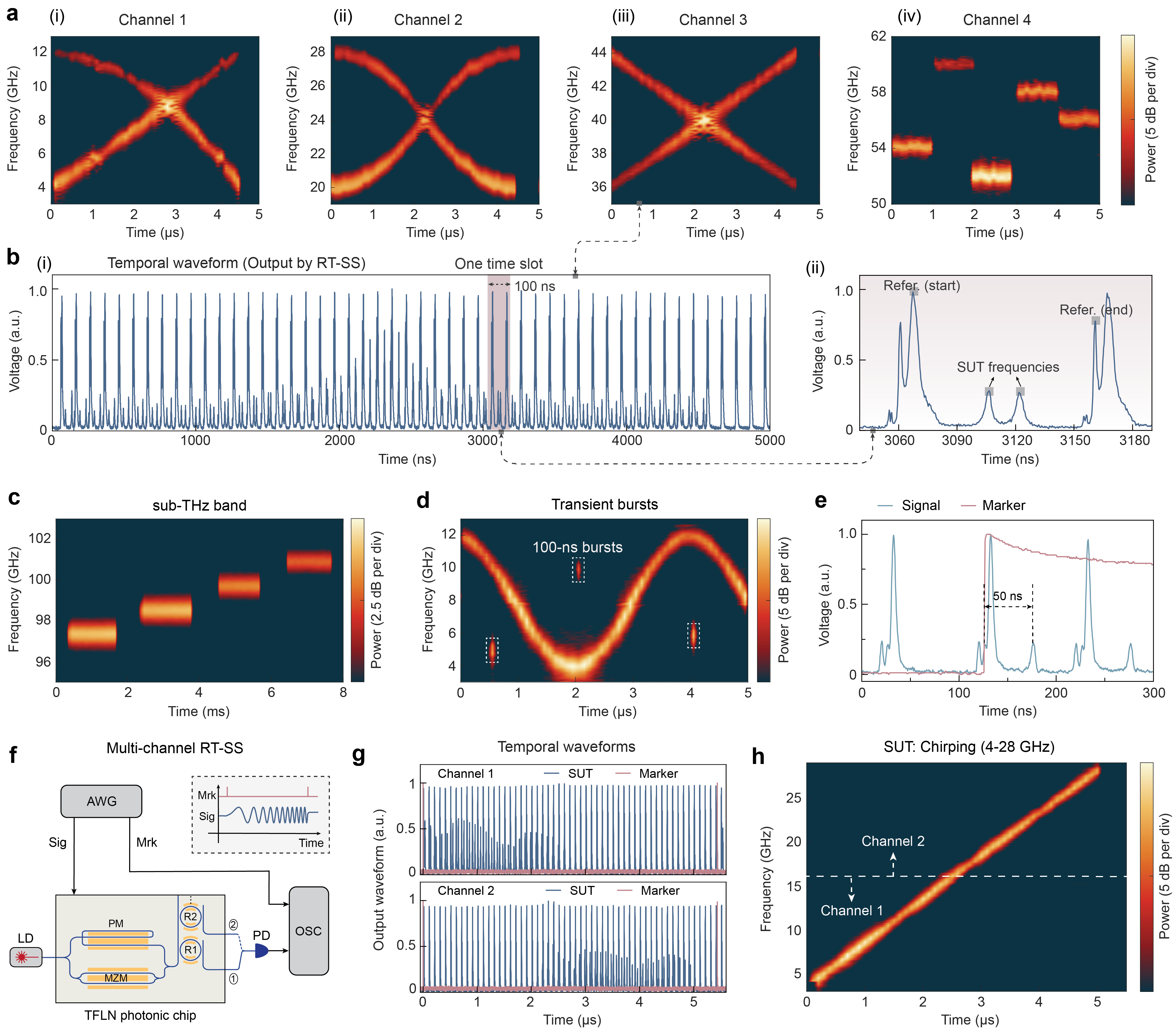}
\caption{\textbf{Full-band photonic RT-SS on frequency-agile RF signals.} \textbf{a}, Reconstructed RF spectrograms of various input signals, including (i) composite linear and quadratic chirp signals (4-12 GHz, channel 1), (ii) double quadratically chirped signals (20-28 GHz, channel 2), (iii) double linearly chirped signals (36-44 GHz, channel 3), and (iv) hopping frequency signals (52-60 GHz, channel 4). \textbf{b}, (i) Original temporal waveform recorded at the output of channel 3 in (\textbf{a}). (ii) Zoom-in view of the temporal waveform in an individual 100-ns time slot. \textbf{c}, Reconstructed spectrogram of a frequency stepping signal near 100 GHz frequency. \textbf{d}, Reconstructed spectrogram of SUT with three transient bursts of 100-ns duration. \textbf{e}, Measured processing latency of the photonic RT-SS system, using an on-off keying 8-GHz sine wave signal as input SUT. The red 'Marker' trace is synchronized with the SUT to provide a time stamp at the burst of the 8-GHz SUT. \textbf{f}, Schematic diagram of the parallel RT-SS setup for measuring input SUTs with ultrawide chirping frequency range across channel 1 and channel 2. AWG, arbitrary waveform generator; LD, laser diode; OSC, oscilloscope. \textbf{g}, Measured temporal waveforms of the RT-SS system output (blue) in channel 1 (top) and channel 2 (bottom), along with the Marker sequences (red) for temporal alignment. \textbf{h}, Seamlessly stitched dual-channel spectrogram for a linearly chirped SUT input in the range of 4-28 GHz. }
\label{fig3}
\end{figure*}

\vspace{3pt}
\noindent\textbf{Ultrawideband real-time spectrum analysis}\\
\noindent To showcase ultrawideband and fast spectrum reconstruction capabilities beyond single-tone frequency identification, we input various types of frequency-agile SUTs and perform joint time-frequency 2D spectrogram analysis. In these experimental RT-SS demonstrations, the scanning speed of the microring filters is further increased to 10 MHz, which corresponds to an analysis time slot of 100 ns, which is 4 orders of magnitude faster than previous RT-SS demonstrations using Si photonics \cite{35,36} thanks to the fast EO tuning of TFLN platform. After re-calibrating the linear FTTM functions of the four spectral channels at the increased scanning speed (see in Supplementary Note V), various SUTs are sent to the photonic RT-SS system, including linear-quadratic composite chirp, dual quadratic chirp, dual linear chirp and frequency-hopping signals. The SUTs are configured with a total signal duration of 5 $\mu$s, which is only limited by the available memory of the arbitrary waveform generator (AWG) used for signal generation. Our proposed photonic RT-SS scheme does not impose any fundamental limitation on the measurement duration and could in principle perform uninterrupted monitoring of incoming RF signals. The output temporal waveforms from the photodetector are sampled by a low-sampling-rate (5 GSa/s) oscilloscope to recover the time-frequency power distributions (spectrograms) of the input SUTs, as shown in Fig. \ref{fig3}a. Comparison with numerically calculated spectrograms of the corresponding input SUTs (Supplementary Note VI) show that the joint time-frequency information of the four different types of SUTs are all recovered with high fidelity, demonstrating the effectiveness and versatility of our TFLN photonic RT-SS system. To provide more detailed information on the spectrogram reconstruction process, we plot the original measured voltage trace of channel 3 in Fig. \ref{fig3}b as an example. It can be seen that the reference pulses appear every 100 ns acting as labels in the temporal domain to locate each analysis time slot. Based on these reference labels, the entire 5-$\mu$s temporal trace is divided into 50 individual 100-ns time slots. Fig. \ref{fig3}b(ii) displays a zoom-in view of a single analysis time slot. It can be observed that, besides the reference pulse at the beginning, a second, smaller reference pulse appears at the end of the scanning period due to the finite return-to-zero time of the actual ramp signal. Nevertheless, the second reference pulse does not affect the spectrogram analysis process, as it falls outside the concerned temporal region in each time slot.

We further perform time-frequency analysis for SUT in the sub-THz spectral range, as shown in Fig. \ref{fig3}c. Here, a stepped frequency signal near 100 GHz is analyzed by the modified sub-THz photonic RT-SS scheme discussed above (Methods and Supplementary Note III for details). To the best of our knowledge, this is the first experimental demonstration of RT-SS in the sub-THz frequency band, which is essential for full-band and dynamic spectral resource management in the 6G era.

In the proposed photonic RT-SS scheme, the Fourier transform (FT) operation is equivalently performed once every 100 ns, which corresponds to $1\times10^{7}$ FTs per second. This 100-ns time slot also determines the temporal resolution of the RT-SS operation. To confirm this, a SUT containing three single-tone bursts of 100 ns duration is generated as input, all of which are successfully captured in the recovered spectrograms (Fig. \ref{fig3}d). Although the Pockels effect on TFLN platform can allow a much higher scan speed, a rapid deterioration in terms of frequency resolution is observed when we further shorten the analysis time slot (see in Supplementary Note VII for more details). This phenomenon is caused by the well-known ringing effect in high-$Q$ optical resonators \cite{50,51}. When a high-$Q$ resonator is excited by a fast frequency-chirping laser source, whose frequency is linearly swept across the resonance with a duration shorter than the cavity lifetime, an oscillating transmission spectrum can be observed leading to broadened temporal pulses \cite{50}. This physical process also happens when a high-$Q$ resonator is fast EO modulated in our case. Therefore, to maintain a reasonable frequency resolution, the temporal resolution is ultimately set at 100 ns in our RT-SS experiments. The achieved temporal resolution significantly outperforms current electronic solutions and meets the requirements in most real-world applications including dynamic spectrum sharing in 6G ISAC scenarios \cite{52}. While photonic RT-SS systems based on the time-lens approach could achieve better temporal resolution (several nano-seconds), they typically suffer from large latencies ($>$100 $\mu$s) \cite{23} and significant bulkiness due to the use of long dispersive fibers, which is completely eliminated in our scheme. In principle, the measurement latency in our scheme is no more than one filter scanning period of 100 ns. To evaluate the latency performance, we simultaneously generate an on-off keying (denoted as ‘1’ and ‘0’ in Fig. \ref{fig3}e) 8-GHz sine wave signal, which is sent to the photonic RT-SS system, and a marker sequence for referencing the time when the signal is turned on. The output temporal waveform from the RT-SS and the marker are collected and analyzed by the same oscilloscope. To minimize the latency, in this experiment, we remove the EDFA (and thus its long internal fiber) and use a high-sensitivity PD (Methods). As shown in Fig. \ref{fig3}e, the measured RT-SS processing latency is $\sim$50 ns in this particular case, which could be shorter or longer (but not more than 100 ns) depending on the relative time of signal appearance within one scanning period.

We further demonstrate the broadband parallel RT-SS ability across multiple channels through a proof-of-principle spectrogram measurement of a linear chirp signal that spans from 4 GHz to 28 GHz using both channel 1 and channel 2 (Fig. \ref{fig3}f). Due to unmatched pitches between the output waveguides and commercial fiber array, the current TFLN chip does not allow capturing output signals from both channels simultaneously. To perform a faithful characterization of both channels, we generate a marker sequence along with the SUT, which temporally labels the beginning of the SUT, as shown in Fig. \ref{fig3}f (details can be found in Method). As a result, separately measured voltage traces of channel 1 and channel 2 can be aligned in the time axis using the marker sequence, as given in Fig. \ref{fig3}g. It can be observed that the two channels capture the time-varying frequency information in the first and second halves of the SUT duration, with partial overlap in the middle to achieve spectral gapless analysis. The combined dual-channel spectrogram is shown in Fig. \ref{fig3}h, clearly reconstructing the entire spectrogram of the SUT without observable discontinuity in the frequency or time axis. Further optimizing the chip layout in future tape-outs will allow us to perform truly parallel RT-SS operations across multiple analysis channels.

\begin{figure*}[ht]
\centering
\includegraphics[width = 18cm]{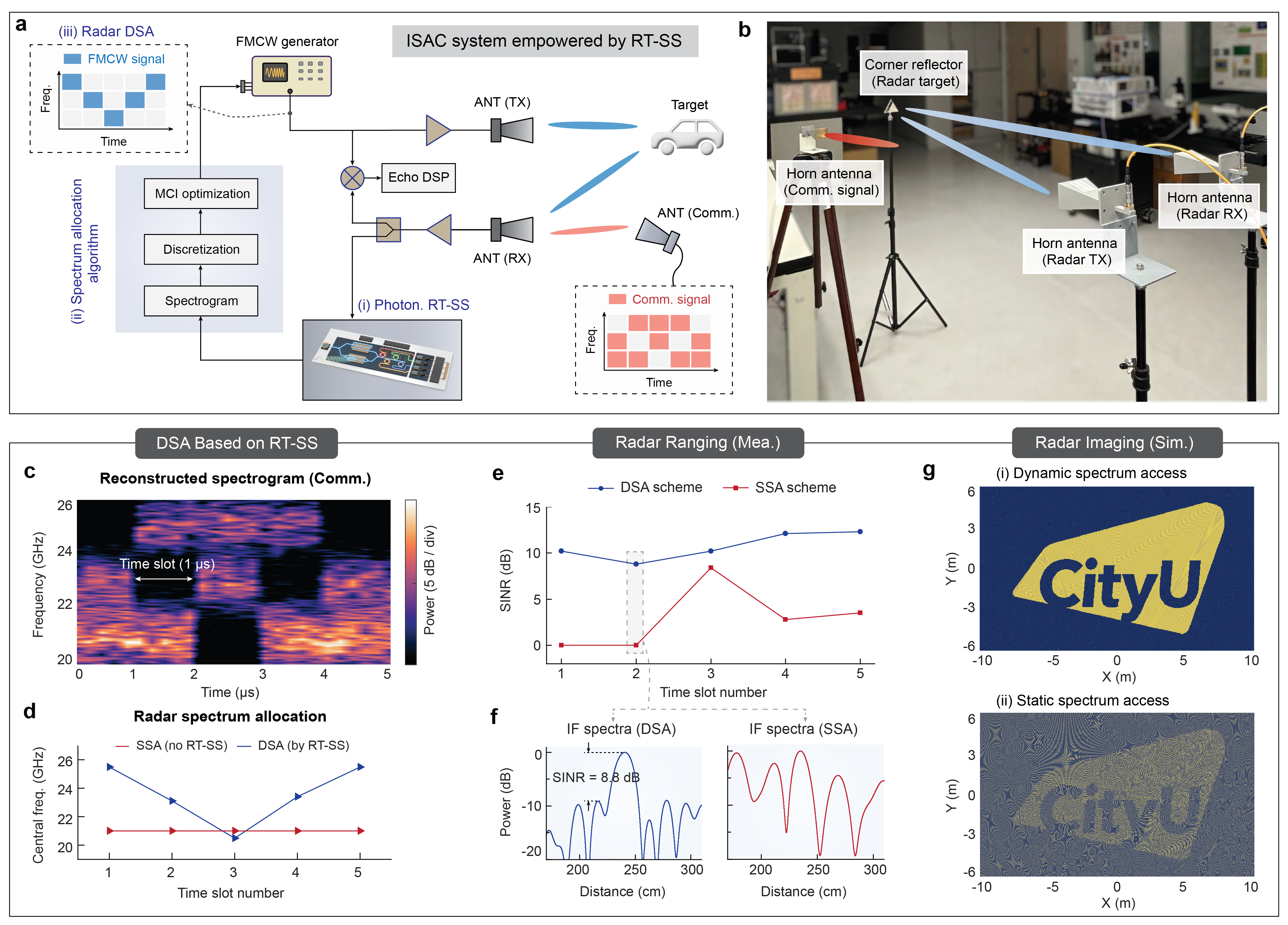}
\caption{\textbf{Photonic RT-SS empowered proof-of-concept ISAC.} \textbf{a}, Illustration of the ISAC experimental setup, where a radar performs target sensing operation based on dynamic spectrum sharing with a communication system emitting frequency-varying data streams. The radar sequentially executes (i) photonic RT-SS, (ii) algorithm-optimized spectrum allocation, and (iii) dynamic hardware spectrum access to coexist with communication. MCI, minimum communication interference; ANT, antenna; FMCW, frequency modulation continuous wave. \textbf{b}, Image of the ISAC experimental setup, where two same horn antennas transmit FMCW waveforms and receive echoes from a corner reflector target. A third horn antenna emits communication signals, acting as spectral interference for radar. \textbf{c}, Reconstructed time-frequency spectrogram of the communication signal by the photonic RT-SS. \textbf{d}, Allocated radar center frequencies in different time slots under the RT-SS-guided DSA and the traditional SSA schemes, both considering a total radar signal bandwidth of 1 GHz. \textbf{e}, Extracted SINRs from the de-chirped intermediate-frequency (IF) electrical spectra of radar echoes, under the DSA and SSA schemes. \textbf{f}, De-chirped IF spectra of the radar echo signals in the second radar time slot in (\textbf{e}). Horizontal axes have been converted from Fourier frequency into distance values for better visualization. \textbf{g}, Comparison of the simulated 2D radar imaging results under the (i) DSA scheme and (ii) SSA scheme. }
\label{fig4}
\end{figure*}

\vspace{3pt}
\noindent\textbf{ISAC demonstration empowered by photonic RT-SS}\\
\noindent Building on the ultrawideband and high-speed spectral analysis capability, we finally deploy the TFLN photonic RT-SS in a proof-of-concept ISAC scenario to demonstrate its real-world application potential. As schematically illustrated in Fig. \ref{fig4}a, we demonstrate that data communication and radar sensing functionalities can operate simultaneously through DSA implementation within a shared spectral resource pool (20-26 GHz), facilitated by the photonic RT-SS. On the communication side, acting as the primary user in this ISAC system, quadrature phase shift keying (QPSK) modulation signals with time-varying spectrum occupancy are generated and emitted into free space by a horn antenna, as shown in Fig. \ref{fig4}b. The spectrum occupancy of the communication signal dynamically varies in time slot of 1-$\mu$s length. On the radar sensing side (secondary user), in addition to the typical transmitting and receiving hardware, the radar is also equipped with photonic RT-SS to capture the spectrum usage of communication activities, thereby enabling its dynamic access to the underutilized spectral regions. Specifically, in each coherent processing interval (CPI), the radar first implements photonic RT-SS to acquire the spectral usage information of the communication channel. We develop a heuristic spectro-temporal resource allocation algorithm (Methods) to adaptively select the optimal frequency bands for radar access, as shown in Fig. \ref{fig4}a [inset (ii)]. In particular, the reconstructed spectrogram after photonic RT-SS is first discretized into multiple small-size spectral resource blocks. Then, the resource block group (comprising several blocks to satisfy radar bandwidth requirement) with minimal communication interference power will be identified through a brute-force search process. This proposed algorithm is versatile for arbitrary temporal resolution, frequency resolution and radar sensing bandwidth budgets. Finally, the radar adjusts its operating frequency range to the algorithm-optimized spectral bands, in which it transmits frequency-modulated continuous wave (FMCW) waveforms to perform ranging measurement of a corner reflector target. In the experiment, the radar CPI is considered as equal to the time-slot length of the communication signal (1 $\mu$s), with one CPI containing a single FMCW waveform. More details about the ISAC experimental setup can be found in the Methods section.

\begin{table*}[htbp]
	\centering
    \renewcommand{\arraystretch}{1.5} 
	\caption{Performance comparison of the state-of-the-art electronic and photonic RT-SS systems}
	\begin{tabular}{p{1.8cm}p{2.4cm}<{\centering}p{2.4cm}<{\centering}p{1.8cm}<{\centering}p{2.0cm}<{\centering}p{2.0cm}<{\centering}p{2.8cm}<{\centering}p{1.4cm}<{\centering}}
	
		\toprule  
		\makecell[c] {Technical \\ scheme} 
        & \makecell[c] {Integration/ \\ Platform}
        & \makecell[c] {Analysis range \\ (GHz)} 
        & \makecell[c] {Latency \\ ($\mu$s)}  
        & \makecell[c] {Temporal \\ resolution \\ (ns)} 
        & \makecell[c] {Frequency \\ resolution \\ (MHz)} 
        & \makecell[c] {Demand on \\ high-speed $^{3}$ \\ electron. devices}
        & \makecell[c] {ISAC \\ demo.} \\
		\midrule 

Electronics\cite{18} & Yes/Silicon  & 3.1-10.6 & $<$420 & N/A $^{2}$ & 132 &  No  & No \\ 

Electronics\cite{19} & Yes/Silicon  & 2-22 & N/A & N/A & 2300 &  No  & No \\ 

Photonics\cite{23} & No  & 0-4.86 & $>$200 & 5 & 340 &  Yes  & No \\

Photonics\cite{29} & No  & 0-0.53 & $<$0.2 & 30 & 30 &  Yes  & No \\

Photonics\cite{33} & No  & 1-9.7 & N/A & 100 & 100 &  Yes  & No \\

Photonics\cite{32} & No  & 0-12 & N/A & 500 & 60 &  Yes  & No \\

Photonics\cite{24} & No  & 2-40 & $>$1$\times$10$^{3}$ & 2 & 2000 &  Yes  & No \\

Photonics\cite{28} & No  & 0-46 & $<$0.1 & 9 & 400 &  Yes  & No \\

Photonics\cite{34} & No  & 0-50.8 & N/A & 500 & 20 &  Yes  & No \\

Photonics\cite{35} & Yes/SOI $^{1}$  & 2-35 & $<$1$\times$10$^{3}$ & 1$\times$10$^{6}$ & 15 &  No  & No \\

Photonics\cite{36} & Yes/SOI  & 1-30 & $<$1$\times$10$^{4}$ & 1$\times$10$^{7}$ & 375 &  No  & No \\

\makecell[l] {Photonics\\ (This work)} & \textbf{Yes/TFLN}  & \textbf{0-62/90-120} & \textbf{$<$0.1} & \textbf{100} & \textbf{350} &  \textbf{No}  & \textbf{Yes} \\

    \bottomrule  
	\end{tabular}
    \begin{tablenotes}
      \item$^{1}$ SOI, silicon on insulator. $^{2}$ N/A, Information not available or not applicable. $^{3}$ High-speed ($>$1 GHz bandwidth) analog-to-digital or digital-to-analog converters used to constitute the RT-SS systems, excluding the equipment for generating the broadband signal under test. 
  \end{tablenotes}
\label{tbl1}
\end{table*}

Fig. \ref{fig4}c shows the reconstructed RF spectrogram of the communication signal, clearly revealing its time-varying spectral activities. Adopting the spectrum allocation algorithm, the optimal frequency ranges for radar dynamic access are identified in every time slot, as shown in Fig. \ref{fig4}d. Here, the blue line represents the algorithm-optimized center frequencies for the proposed DSA scheme, while the red line represents the traditional SSA scheme (as comparison baseline), both considering a radar sensing bandwidth of 1 GHz. It is worth noting that the sensing bandwidth does not need to be fixed at 1 GHz in the DSA implementation and can be flexibly adjusted according to practical resolution requirements. The de-chirped radar echoes of each time slot in both spectrum access schemes are collected. After a fast Fourier transform process, the electrical spectra of the echoes can be obtained. The SINR of the echo spectra is extracted for all time slots and plotted in Fig. \ref{fig4}e. In the DSA scheme, the measured SINR values are all higher than 8.8 dB, indicating high-quality target ranging measurements. On the contrary, the SINR is severely deteriorated in the SSA scheme, except for the third time slot which is not spectrally interfering with communication. The comparative ranging results confirm the effectiveness of the RT-SS-guided DSA implementation in maintaining high-performance radar operations in the presence of congested communication interferences.

In the 6G ISAC vision, the radar is also expected to perform high-resolution imaging tasks in emerging scenes such as autonomous driving, smart medical system, and intelligent manufacturing. To validate the RT-SS-enabled DSA performance in imaging tasks, we simulate 2D radar imaging results of a target with 'CityU' logo based on the DSA and the SSA schemes, respectively, using the same communication signals and radar waveforms as those in the ranging measurement. The DSA and SSA imaging results in Fig. \ref{fig4}g exhibit stark contrast in imaging quality, confirming the effectiveness of photonic RT-SS-based radar dynamic access in alleviating time-varying communication spectral interferences. Based on these proof-of-principle ranging and imaging demonstrations, we believe our TFLN photonic RT-SS holds significant potential for high-efficiency dynamic sharing of scarce spectral resources in future 6G ISAC networks.

\noindent
\textbf{Discussion} \\
Leveraging the large EO modulation bandwidth and fast filter tuning speed empowered by the Pockels effect, along with a parallel spectral measurement scheme referenced by an EO comb, our photonic chip-based RT-SS system achieves significantly improved overall performances, especially in spectral analysis bandwidth, integration level, and processing latency, compared to previous electronic and photonic RT-SS demonstrations, as shown in Table \ref{tbl1}. These unique features are particularly important for ISAC applications in future 6G wireless networks, enabling effective utilization and dynamic management of the full spectrum resources from microwave to sub-terahertz with significantly reduced SWaP. Building on the excellent scalability of the TFLN platform, the proposed RT-SS architecture could be readily extended deeper into the THz band (e.g. 100 – 300 GHz) leveraging ultrabroadband TFLN modulators \cite{53}, a larger-scale microring filter bank, and a reasonably broad EO comb generator \cite{42}. The spectral resolution of our photonic RT-SS system could be further improved by adopting microring filters with narrower linewidths. Recently, TFLN microrings with ultrahigh $Q$ factors exceeding 29 million have been achieved through high-quality dry etching and wider waveguide designs \cite{54}. By employing such a high-$Q$ microring filter, a frequency resolution down to 10 MHz is possible, which however would come at the cost of degraded temporal resolution as discussed above.

Moving forward, the TFLN photonic RT-SS core could be further integrated and/or co-packaged with the other peripheral photonic and electronic components toward compact and low-cost real-world deployment in ISAC base stations. The laser source and PD array can be integrated into the TFLN platform using hybrid or heterogeneous integration technologies \cite{55}, potentially reducing optical coupling loss and eliminating the need for the optical amplifier used in the present work. All electronic peripherals in our scheme are implemented using low-speed circuitry, allowing for straightforward hybrid integration or co-packaging with photonic circuits at a low cost. More importantly, benefiting from the versatility and scalability of the TFLN photonic platform, the RT-SS module developed in this work can be seamlessly combined with other functional microwave photonic modules including transmitter \cite{56}, receiver \cite{57}, and analog signal processor \cite{47}, towards a photonics-driven wireless base station featuring full-spectrum utilization, ultralow SWaP, and dynamic adaptability to complex electromagnetic environment in the 6G era.

\vspace{6pt}
\noindent \textbf{Methods}\\
\begin{footnotesize}
\noindent \textbf{Design and fabrication of the TFLN photonic circuit} 

\noindent The on-chip phase modulator and MZ modulator are designed with an effective modulation length of 1 cm. The widths of the signal and ground metal strips are 50 $\mu$m and 250 $\mu$m, respectively. Slot-electrode structure is utilized and simulated using High Frequency Simulation Software (Ansys HFSS) to achieve velocity matching between lightwave and microwave, leading to a high EO bandwidth. The microring resonators are designed in a racetrack configuration, with a waveguide width of 2 $\mu$m and an electrode gap of 6 $\mu$m to reduce the optical propagation loss. To suppress high-order mode excitation while minimizing the bending radius, an Euler curve is utilized and simulated using Ansys Lumerical FDTD to ensure an adiabatic mode transition in the bending region. The effective radius of the Euler bend is 80 $\mu$m, and the length of the straight region is 180 $\mu$m, yielding a free spectral range (FSR) of approximately 125 GHz. For the assessment of frequency resolution, a larger racetrack microring is used, with its straight region increased to 600 $\mu$m to obtain a higher loaded $Q$ factor, leading to a FSR of 70 GHz.

The devices are fabricated using a commercially available x-cut LNOI 4-inch wafer (NANOLN), with a 500 nm LN thin film, a 4.7 $\mu$m buried SiO$_{2}$ layer and a 500 $\mu$m silicon substrate. First, SiO$_{2}$ is deposited onto the surface of the LN wafer as an etching hard mask using plasma-enhanced chemical vapor deposition (PECVD). The optical waveguides and passive devices are then patterned onto the wafer using ultraviolet stepper lithography. Next, the exposed resist patterns are first transferred to the SiO$_{2}$ layer using a standard fluorine-based dry etching process, and then to the LN device layer using an optimized Ar$^{+}$-based inductively coupled plasma reactive-ion etching process. The LN thin film is etched with a depth of about 250 nm, leaving a 250-nm-thick slab layer. After removal of the residual SiO$_{2}$ mask and redeposition, the sample is annealed. Then, the RF electrodes and wires/pads for electrical connections are severally fabricated through a second, third, and fourth lithography and metal lift-off processes. Finally, the TFLN chips are carefully cleaved for edge optical coupling with a coupling loss of approximately 4-5 dB per facet.

\vspace{6pt}
\noindent\textbf{Characterization of on-chip building blocks} 

\noindent For the on-chip MZM, the EO modulation bandwidth is evaluated by the small-signal EO S21 response, which is measured by a vector network analyzer (VNA, Keysight N5227B). The frequency-sweeping RF signal is applied onto the RF electrodes through a broadband probe (GGB Industries, 67 GHz). The modulated optical signals are captured by a high-speed photodetector (Finisar XPDV412xR, 100 GHz) and then sent back to the VNA. The RF cable losses, probe loss and photodetector response are calibrated and de-embedded from the measured S21 responses. To measure the half-wave voltage, a kilohertz electrical triangular waveform is generated from a function generator (Rigol DG4162) to drive the device. The output optical signal of the MZM is detected using a low-speed PD (New Focus 1811) and monitored using an oscilloscope (Rigol DS6104). To characterize the EO comb generation, the phase modulator is driven by a tunable microwave source with an RF power of 27 dBm. The optical spectra are measured by an optical spectrum analyzer (OSA, Yokogawa AQ6370D). Due to the recycling configuration, the phase modulation efficiency is maximized at periodically appearing optimal RF frequencies, when the recycled optical signal remains in phase with the driving microwave signal. In this work, the EO comb spectra at a 32-GHz repetition rate are characterized.

To test the optical performances of the microring filter bank, a wavelength-tunable laser source (Santec TSL-570) of 13 dBm power is sent to the devices using a lensed fiber (mode size of 2 $\mu$m diameter), and the optical signals from the drop ports of the microrings are collected using a second lensed fiber and sent to a photodetector (New Focus 1811). The recovered electrical signals are recorded by a data acquisition module (NI USB-6259), which are synchronized with the output trigger of the laser source. In this way, the transmission spectra can be obtained in a high wavelength resolution of 0.1 pm. To measure the EO scanning range of the microring, a broadband spontaneous emission noise source from an EDFA (Amonics) is injected into the device, and the output optical signals at the drop port of the microring filters are collected using an OSA. The scanning period of the microring is set as 1 $\mu$s, which is much shorter than a single-cycle measurement time (second-level) of the OSA. Hence, a flattop transmission spectrum is obtained, which corresponds to the frequency sweeping range of the microring filter.

\vspace{6pt}
\noindent\textbf{Experimental setup of TFLN photonic RT-SS} 

\noindent In the photonic RT-SS experiments, a c.w. optical carrier of 13 dBm power is generated by a tunable laser source (Santec TSL-710) and then coupled into the TFLN chip through a lensed fiber. A polarization controller is adopted to ensure TE polarization of the injection light for the largest EO modulation efficiency. For the generation of reference EO comb, the recycling PM is driven by a 32 GHz sinusoidal wave, which is provided by a microwave signal generator (Agilent E8244A) and then amplified by an electrical amplifier (Plugtech EVK-RF-A-40-G). The input signal under test (SUT) in the frequency range of 0-32 GHz is directly generated by a 92-GSa/s arbitrary waveform generator (AWG, Keysight 8196A). In the frequency range of 32-64 GHz, the SUT is first generated within 0-14 GHz by the AWG and then upconverted through a mixer (14-65 GHz) and an RF local oscillator (R\&S SMA 100B). The generated SUT is further amplified by an electrical low-noise amplifier (SHF 807C) and combined with a DC bias voltage with the help of a broadband bias-tee (SHF BT-65). A five-pin GSGSG RF probe (GGB Industries, 50 GHz) is employed to apply the 32 GHz local oscillation and the broadband SUT onto the metal pads simultaneously.

A function generator (500 MSa/s, Rigol DG4162) is employed to apply a 20 Vpp ramp voltage waveform to the EO-scanning microring filter bank. The output optical signals by the TFLN chip containing desired spectral information of SUT, are collected by another lensed fiber. The optical signals are amplified by an EDFA (Amonics) to compensate the fiber coupling loss. A bandpass optical filter (Santec OTF-930) is used after the EDFA to suppress the out-of-band spontaneous emission noise. The optical signals are finally converted back to electrical domain by a PD with internal trans-impedance amplifier (12 GHz, New Focus 1544), and captured by a low-speed oscilloscope (5 GSa/s, R\&S MXO44-245). The PD is reused for serial detection across four channels. The collected temporal electrical waveforms from the oscilloscope are off-line processed in a desktop to obtain the joint time-frequency spectrogram. In the latency measurement, the EDFA is removed from the setup due to its significant contribution to latency from the long internal fiber (85 m). In this case, to ensure effective output signal detection, a photodetector (New Focus 1811, 125 MHz) with higher sensitivity is employed with a compromise in bandwidth.

To measure the SUT across two spectral analysis channels (channel 1 and channel 2), the Marker function of the AWG (Keysight 8196A) is adopted, which can provide a user-defined binary sequence that is synchronized in time with the SUT stream. Here we code the marker signal at a high level at the beginning of the SUT duration, while at other times it is at a zero level. In the experiment, the marker and the photo-detected SUT signals are collected by the two channels of the oscilloscope simultaneously. Using the markers, the electrical temporal signals produced from the two channels can be aligned on the time axis and seamlessly stitched together to reconstruct the complete spectrogram of the SUT, covering its entire frequency range. In the system latency measurement, the Marker function is also employed to label the instantaneous frequency information of input SUT in time axis, so that the time interval between SUT input and the RT-SS output can be evaluated.

\vspace{6pt}
\noindent\textbf{Photonic RT-SS in sub-THz band} 

\noindent In the experimental demonstrations of TFLN photonic RT-SS in the sub-THz frequency range (90-120 GHz), a c.w. optical carrier is generated by a laser source (Santec TSL-710) to modulate the SUT onto the optical domain. A second laser reference is produced by another source (Santec TSL-510). The carrier and reference light waves are coupled into the TFLN chip using a dual-channel lensed fiber array. The wavelength of the reference laser source was tuned to achieve various frequency spacing between the carrier laser, for supporting the frequency measurement in different spectral range. The SUT is first generated by an RF local oscillator (R\&S SMA 100B) and then undergoes eight-fold frequency multiplication by an electrical multiplier (90-140 GHz, Millitech AMC-08-RFH00), delivering an output RF power of 4 dBm. The output signal from the multiplier is delivered to the on-chip electrical contact pads through a metallic waveguide and a high-speed three-pin GSG RF probe (GGB Industries, 90-140 GHz). The on-chip MZM device is designed with unbalanced optical paths, therefore, the null bias-point can be realized though adjusting the optical carrier wavelength. Considering that the modulated sidebands in the 100 GHz band have weaker power than those in the relatively low frequency range ($<$62 GHz), a PD with higher sensitivity but lower bandwidth (125 MHz, New Focus 1811) is employed to detect the output signals from the TFLN photonic chip. The periodic scan frequency of the microring filter is configured within the 10 kHz to 100 kHz range to ensure that the generated temporal pulses do not exceed the operating bandwidth of the PD. For the generation of the stepped frequency signal, the frequency sweeping function of the microwave source (R\&S SMA 100B) is adopted. The dwell time is set as 2 ms, which is the minimum value supported by the instrument.

\vspace{6pt}
\noindent\textbf{ISAC demonstration enabled by photonic RT-SS} 

\noindent In the proof-of-concept ISAC experiments, radar sensing and communication signals dynamically share the same spectral resource pool in the K-band (20-26 GHz), which is limited by the available RF equipment and devices rather than the bandwidth of the proposed photonic RT-SS. The entire ISAC setup consists of three main parts: the communication hardware, the radar hardware, and the photonic RT-SS module attached to the radar to guide its dynamic spectrum access. For the communication (primary user) hardware, a series of QPSK modulation signals is generated by an AWG (Keysight 8196A), featuring time-varying spectral occupancy with a time slot length of 1 $\mu$s and a total signal duration of 5 $\mu$s. 

Specifically, in each time slot, two independent streams of 1.5 Gbaud/s QPSK-format baseband data are modulated on two K-band RF carriers. From the 1st to the 5th time slot, the dual carrier frequencies are as follows: 21 GHz and 23 GHz (time slot 1), 21 GHz and 25 GHz (time slot 2), 23 GHz and 25 GHz (time slot 3), 21 GHz and 25 GHz (time slot 4), and 21 GHz and 23 GHz (time slot 5). The generated communication signal is amplified by an electrical amplifier (Pasternack PE15A4021) and then emitted to free space through a horn antenna. 

For the radar (secondary user) hardware, a corner reflector is employed as the ranging target. In the radar transmitter, FMCW waveforms are generated using another channel of the AWG, which are amplified by an electrical amplifier (Pasternack PE15A5054) and then emitted to free space through a horn antenna. In the radar receiver, the echoes are collected by another horn antenna of the same type as the transmitter and amplified by a low-noise amplifier (SHF S807). Then, the echoes are mixed with the branched-out transmitted waveform to perform de-chirping operation. The de-chirped radar echo is captured by an oscilloscope (R\&S MXO44-245). Finally, the time-domain echo signals are processed off-line by a desktop, where an FFT process is executed to transform the echo signals from temporal domain to frequency domain. Based on the echo spectra, the distance of the radar target can be calculated. 

The photonic RT-SS module is attached to the radar hardware, to guide the radar in dynamically accessing the underutilized spectral regions by communication. The receiving antenna of radar is used in parallel for photonic RT-SS. The collected signals from free-space are connected to the TFLN photonic RT-SS chip using RF cables and a high-speed probe. In a real-world ISAC network, the radar sensing operation and RT-SS operation are supposed to execute synchronously for the DSA implementation. This demands the radar transmitting hardware to have high-speed waveform reconfigurability with latency shorter than the spectral variation time of the communication sequence. However, the AWG in our setup requires second-level time to reload a new FMCW waveform. Therefore, in the ISAC system demonstration, we conduct the experiment in a quasi-real-time approach. Specifically, we first perform the photonic RT-SS to completely record the time-varying spectral activities of communication signal over its total duration consisting of 5 time slots (5 $\mu$s). Then, we use the dynamic spectrum allocation algorithm to search the optimal underutilized spectral regions by communication in every time slot, conducted offline in a desktop. Finally, the radar adopts the algorithm-optimized frequency ranges to dynamically transmit FMCW waveform for target ranging, to coexist with the communication sequence over its entire duration (5 $\mu$s). The radar sensing and communication signals are synchronized in time, since they are generated by two output channels of the same AWG. 

In addition to the ranging measurement, target imaging functionality is also necessary for a radar in 6G ISAC networks. Therefore, we also compare the radar imaging performances in this spectrum-shared ISAC scenario based on simulation, under the RT-SS-guided DSA and the traditional SSA schemes, respectively. In the radar imaging simulation, the transmitting and receiving antennas are both positioned at [0, 0] meters, assuming it has no actual physical dimensions. A complex two-dimensional radar sensing target is constructed using geometric patterns derived from a binarized image of the 'CityU' logo, which is mapped onto a spatial grid spanning [-10, 10] meters. The synthesized FMCW signals in both the DSA and SSA schemes, as well as the communication signals (as interference for radar) generated in simulation are kept same with those in the ranging experiment. To achieve the complete 2D imaging of the target, the antenna is rotated across 360 degrees with discrete steps. At each angle, FMCW sensing waveforms are emitted to the target and the reflection echoes are collected. The received echoes are processed through de-chirping. Finally, the processed echoes from all discrete angles are spatially accumulated to produce a normalized two-dimensional radar image of the 2D target.

\vspace{6pt}
\noindent\textbf{Adaptive spectrum allocation algorithm} 

\noindent The proposed spectro-temporal resource allocation algorithm operates over discrete time frames of the DSA-based ISAC system, where communication is considered as the primary task, and radar sensing is the secondary task. In each time frame, RT-SS is first performed to acquire a sequence of discretized spectrum occupancy data over the entire shared spectral range of the ISAC system, revealing the interference power from communication. Here, let $R(f)$ denotes the measured spectral power sequence by the RT-SS operation, $\Delta f$ denotes the sampled frequency step, and $M$ denotes the total number of frequency samples obtained in a time frame. In this way, the discretized frequency samples are given by:

\begin{equation}
f = \left\{ {{f_{\min }}, \ldots ,{f_{\min }} + \left( {m - 1} \right)\Delta f, \ldots ,{f_{\min }} + \left( {M - 1} \right)\Delta f} \right\}
\label{EqS1}
\end{equation}

\noindent Suppose the required radar sensing bandwidth is ${B_s}$, which is usually determined by the needed sensing resolution in practical applications. The corresponding number of frequency samples required for the radar sensing is then determined as follows:

\begin{equation}
{M_s} = \left\lceil {\frac{{{B_s}M}}{B}} \right\rceil 
\label{EqS2}
\end{equation}

\noindent where the $B$ denotes the total bandwidth of the shared spectral range as ${f_{\max }} - {f_{\min }}$. Next, the algorithm evaluates the average interference power level over all potential radar operating frequency bands using:

\begin{equation}
{R_m} = \frac{1}{{{M_s}}}\sum\limits_{{f_{\min }} + \left( {m - 1} \right)\Delta f}^{{f_{\min }} + \left( {{M_s} + m - 2} \right)\Delta f} {R\left( f \right)}
\label{EqS3}
\end{equation}

\noindent where $m \in \left\{ {1, \ldots ,M - {M_s} + 1} \right\}$. Then, the core of the algorithm is to solve the following integer programming problem with a limited search space in order to select the radar frequency band that minimizes the average interference:

\begin{equation}
\begin{array}{l}
\begin{array}{*{20}{c}}
{\mathop {\min }\limits_m }&{{R_m}}
\end{array}\\
\begin{array}{*{20}{c}}
{{\rm{s}}{\rm{.t}}{\rm{.}}}&{x \in \left\{ {1, \ldots ,M - {M_s} + 1} \right\}}
\end{array}
\end{array}
\label{EqS4}
\end{equation}

\noindent Let $m^{*}$ denote the optimal solution to this problem. Note that, due to the constrained variable space, the optimization can be efficiently executed via a brute-force search. Finally, the optimal radar transmission frequency band can be obtained. The central frequency of the allocated radar spectral block is identified as:

\begin{equation}
{F_c} = {f_{\min }} + \frac{{\left( {{M_s} + 2{m^*} - 3} \right)}}{2}\Delta f
\label{EqS5}
\end{equation}

\noindent which corresponds to the minimal average interference level, thereby ensuring effective and low-interference radar sensing operation.

\vspace{10pt}
\noindent\textbf{Data availability}\\
The data that supports the plots within this paper and other findings of this study are available from the corresponding authors upon reasonable request.

\vspace{10pt}
\noindent\textbf{Code availability}\\
The codes that support the findings of this study are available from the corresponding authors upon reasonable request.

\end{footnotesize}
\vspace{20pt}

\bibliography{REF.bib}


\vspace{12pt}
\begin{footnotesize}

\vspace{6pt}
\noindent \textbf{Acknowledgment}

\noindent 
We thank W.-H. Wong and K. Shum at CityU for their help in device fabrication and measurement. We thank the technical support of C. F. Yeung, S. Y. Lao, C. W. Lai and L. Ho at HKUST, Nanosystem Fabrication Facility (NFF), for the stepper lithography and PECVD process. This work is supported by the Research Grants Council, University Grants Committee (CityU 11204022, CityU 11204523, C1002-22Y, STG3/E-704/23-N, CityU 11212721, JRFS2526-1S01), Croucher Foundation (9509005), City University of Hong Kong (9610682).

\vspace{6pt}
\noindent \textbf{Author contributions}

\noindent 
Y.T., H.F. and C.W. conceived the idea, with the discussions from H.S. and X.W. Y.T. proposed the system architecture and designed the devices. H.F., Y.W. and Z.C. fabricated the devices. Y.T. and H.F. carried out the experimental measurements and data analysis, with the assistant from X.X., Y.S.Z., Y.W.Z., T.G, and Z.T. The adaptive spectrum allocation algorithm was proposed by Y.F., X.Y. and J.X. The spectrogram data analysis and the imaging simulation of the ISAC demonstration was implemented by Y.F., with the guidance from X.Y. The paper was prepared by Y.T. and H.F., with contributions from all authors. C.W. supervised the project.

\vspace{6pt}
\noindent
\textbf{Additional information} 

\noindent Supplementary information is available in the online version of the paper. Correspondence and requests for materials should be addressed to Y.T., H.F., and C.W.

\vspace{6pt}
\noindent \textbf{Competing financial interests} 

\noindent H.F., Z. C. and C.W. are involved in developing lithium niobate technologies at RhinoptiX Technology Ltd.
\end{footnotesize}

\end{document}